\documentclass[twocolumn,english,aps,prb,floatfix]{revtex4}
\usepackage[T1]{fontenc}
\usepackage[latin2]{inputenc}
\usepackage{amssymb}

\makeatletter

\usepackage{graphicx}
\usepackage{bm}

\usepackage{babel}
\makeatother
\begin{document}

\title{Shot noise reduction in quantum wires with ''0.7 structure''}

\author{A. Ram\v{s}ak$^{1,2}$ and J. H. Jefferson$^{3}$}

\affiliation{$^{1}$Faculty of Mathematics and Physics, University of Ljubljana,
Slovenia}

\affiliation{$^{2}$J. Stefan Institute, Ljubljana, Slovenia,}

\affiliation{$^{3}$QinetiQ, Sensors and Electronics Division, St. Andrews Road,
Great Malvern, England}

\begin{abstract}
Shot noise reduction in quantum wires is interpreted within the model
for the ''0.7 structure'' in the conductance of near perfect quantum
wires {[}T. Rejec, A. Ram{\v s}ak, and J. H. Jefferson, Phys. Rev.
B \textbf{62}, 12985 (2000){]}. It is shown how the Fano factor structure
is related to the specific structure of the conductance as a consequence
of the singlet--triplet nature of the resonances with the probability
ratio 1:3. An additional feature in the Fano factor, related to the
''0.25 structure'' in conductance, is predicted. 
\end{abstract}


\maketitle
Conductance in various types of quantum wires and quantum point contacts
is quantized in units of $G_{0}=2e^{2}/h$. Since the first experimental
evidence for this effect\cite{wees88,wharam88} many subsequent experiments
have supported the idea of ballistic conductance in clean quantum
wires. However, certain anomalies remain, some of which are believed
to be related to electron-electron interactions and appear to be spin-dependent.
In the rising edge to the first conductance plateau, a structure appears
around $0.7G_{0}$, merging into the plateau at higher energies.\cite{thomas96}
Traces of such an anomaly are present already in the early measurements.\cite{wees88,patel91}
In quantum point contacts an additional structure appears around $0.25G_{0}$
with increasing source-drain bias \cite{thomas96} and this structure
is also seen at low bias in hard-confined quantum wires.\cite{patel91,philmag98,kaufman99,cronenphd,picciotto04} 

Under increasing magnetic field the $0.7$ structure moves down and
merges with the $\frac{1}{2}G_{0}$ plateau at very high fields.\cite{patel91,philmag98,kaufman99,cronenphd,picciotto04}
At elevated temperatures the structures are eventually washed out
but surprisingly this also happens at very low temperatures where
the disappearance of the 0.7 structure signals the formation of a
Kondo-like correlated spin state.\cite{cronenphd,cronenprl} Related
anomalies in thermopower measurements may also be explained though
a violation of the Mott-law emerges at very low temperatures. This
additionally suggests a many-body nature of electron transport in
ballistic quantum wires.\cite{appleyard}

Recently the non-equilibrium current noise in a one-dimensional quantum
wire was measured\cite{will,roche} and this new class of measurements
opens up a range of possibilities for gaining new insight into the
problem of the transport anomalies in clean quantum wires. In particular,
careful measurements of the Fano factor in the expression for shot
noise for this system appears to be correlated with anomalies in conductance,
signaling different transmission probabilities for spin sub-channels.

In this paper we present a theoretical explanation of the peculiar
dependence of the Fano factor on conductance and Fermi energy in ballistic
quantum wires. We follow the idea developed in the study of conductance
of a clean one-dimensional quantum wire with cylindrical\cite{rrj00}
or rectangular cross section\cite{jphys00} and a very weak bulge.
Several other scenarios for the 0.7 structure were recently proposed,
from a phenomenological model involving enhanced spin correlations,\cite{bruus01}
Kondo-like physics due to a localised moment,\cite{meir02} fluctuation
of local electron density\cite{sushkov03} to corrections due to the
backscattering of electrons due to phonons or Wigner crystal state
formation.\cite{matveev03} The relevance of these models to the description
of the 0.7 anomaly in shot noise measurements has not yet been investigated.

Here we do not limit the investigation to a particular geometry of
the wire but consider the general case of a slightly imperfect quantum
wire. Such structures occur naturally, e.g. from a two-dimensional
electron gas (2DEG) in which a surface split-gate, which depletes
the 2DEG below it, gives rise to a quasi one-dimensional conducting
channel at low temperatures. Slight deviation from a perfect one-dimensional
confining potential, either accidental or deliberate, can give rise
to a localized potential well with single-electron bound states. This
occurs in a wire with a weak symmetric bulge\cite{rrj00} or in the
presence of remote gates, impurities or even electric polarization
due to the electron itself. The lowest bound state with energy $\epsilon_{b}<0$
relative to the bottom of the first conductance channel is shown schematically
in Fig.~1(a). Provided the confining potential is sufficiently weak,
only a single electron will be bound since the energy of the second
electron will be in the continuum due to Coulomb repulsion and the
system behaves like an open quantum dot. From the numerically exact
solution of two-electron scattering problem, one can extract the transmission
probabilities for particular spin configurations.\cite{rrj00} The
problem is analogous to treating the collision of an electron with
a hydrogen atom, e.g., as studied by J. R. Oppenheimer and N. F. Mott.\cite{oppenheimer}
This can be visualized as the scattering of the second electron in
an effective potential $V_{eff}$ arising from the combined effect
of the Coulomb repulsion from the first electron and the initial confining
potential {[}Fig.~1(b){]}. In the absence of a magnetic field the
appropriate spin sub-channels are singlet and triplet and, as illustrated
in Fig.~1(b), the two-electron system exhibits singlet and triplet
quasi-bound-state resonances in transmission. Summing over all electrons
in the leads gives the current,\cite{landauer,rrj00}

\begin{equation}
I=\frac{2e}{h}\int(\frac{1}{4}T_{0}+\frac{3}{4}T_{1})(f_{L}-f_{R})d\epsilon,\label{eq:lbg}\end{equation}
 where $T_{S}=T_{S}(\epsilon)$ is energy dependent singlet or triplet
transmission probability for $S=0$ and $S=1$, respectively. $f_{L,R}=\{1+\exp[(\epsilon-\mu_{L,R})/k_{B}T]\}^{-1}$
is the usual Fermi distribution function corresponding to left and
right lead, respectively, with temperature $T$ and the Boltzmann
constant $k_{B}$. In the linear regime, $\Delta\mu=\mu_{L}-\mu_{R}\to0$,
this reduces to a generalized Landauer-B\"{ u}ttiker formula\cite{landauer,rrj00}
for conductance, $G=eI/\Delta\mu$. The many-electron problem can
be mapped onto an extended Anderson model\cite{pecs,rrj03} for which
we have an \emph{open} quantum dot with Coulomb blockade except near
the resonances which are analogue to ''mixed-valence'' single-electron
tunneling regime. At very low temperatures the effects of Kondo physics
are expected and indeed signaled experimentally\cite{cronenprl} and
studied theoretically.\cite{meir02} However, at higher temperatures
these Kondo effects are suppressed and the extended Anderson model
yields a conductance in agreement with Eq.~(\ref{eq:lbg}).

In accordance with the Lieb-Mattis theorem\cite{lieb} the singlet
resonance is always at lower energies than the triplet, $\epsilon_{0}<\epsilon_{1}$,
and consequently the quasi-bound state has longer lifetime (the resonance
is sharper) than the triplet. This is clearly seen from the results
obtained for the case of a cylindrical (or rectangular) quantum wire
with a symmetric bulge,\cite{rrj00,jphys00} presented in Fig~2(${\textrm{a}}_{1}$)
and Fig~2(${\textrm{b}}_{1}$).

Recent high accuracy shot noise measurements enabled the extraction
of the Fano factor in ballistic quantum wires.\cite{roche} The Fano
factor $F$ is a convenient measure of the deviation from Poissonian
shot noise. It is the ratio of the actual shot noise and the Poisson
noise that would be measured in an independent-electron system.\cite{blanter}
This factor is, in our model,

\begin{equation}
F=\frac{\int[T_{0}(1-T_{0})+3T_{1}(1-T_{1})](f_{L}-f_{R})^{2}d\epsilon}{\int(T_{0}+3T_{1})(f_{L}-f_{R})^{2}d\epsilon}.\label{eq:f}\end{equation}
 This expression and Eq.~(\ref{eq:lbg}), are based on the results
of a two-electron scattering between a single bound electron and a
propagating conduction electron with a summation over all conduction
electrons near the Fermi energy. This approximation is only valid
at temperatures above the Kondo scale in this system,\cite{meir02}
as discussed in Ref.~\onlinecite{rrj03}. Eq.~(\ref{eq:f}) directly
reflects the fact that singlet and triplet modes do not mix in this
pairwise interaction approximation, resulting in contributions to
the noise that add incoherently with the probability ratio 1:3 for
singlet and triplet scattering.

The conductance $G(\mu)$ and Fano factor $F(\mu)$ are plotted vs
$\mu$ in Fig.~2(${\textrm{a}}_{2}$) and Fig.~2(${\mathrm{b}}_{2}$),
with Fig.~2(${\textrm{a}}_{3}$) and Fig.~2(${\mathrm{b}}_{3}$)
showing the Fano factor $F$ vs $G$ for various temperatures and
in the linear response regime, $\mu=\mu_{L}\sim\mu_{R}$. The dotted
lines show zero temperature boundaries for the allowed values of $F$,
under the assumption of validity of Eq.~(\ref{eq:f}) in the limit
$T\to0$ and the unitarity condition for the transmission probabilities,
$0\leq T_{S}\leq1$. Eq.~(\ref{eq:f}) is not strictly valid in the
limit $T\to0$ due to many-electron effects which start becoming important
at low temperatures. Thus this zero-temperature limit should be regarded
as a limiting behavior that would occur in the absence of such many-body
effects. The Fano factor exhibits two distinctive features. Firstly,
there is a structure for $G/G_{0}<0.5$ corresponding to the sharp
$0.25$ singlet conductance anomaly. The second distinctive feature
is in the region $0.5<G/G_{0}<1$ and corresponds to the dip in singlet
channel just above the singlet resonance and also partially to the
triplet channel resonance. In our previous work we assumed a symmetric
confining potential fluctuation, giving perfect transmission probabilities
at resonance energies. However, in real systems left-right symmetry
will not be perfect, especially if the fluctuation is of random origin,
and also under finite source-drain bias. In these cases $T_{S}<1$
even on resonance. Such an example is presented in Fig.~2(${\textrm{c}}_{1-3}$).
In this case the structure of $F(G)$ is less pronounced, consisting
of kinks at $G/G_{0}\lesssim0.5$ and $G/G_{0}\lesssim0.75$. This
behavior is a consequence of the absence of a pronounced triplet resonance
and a dip in the singlet channel as mentioned above. Such a situation
is typical for very weak confining potential fluctuations, where the
triplet resonance is far in the continuum. 

The structure at $G/G_{0}\lesssim0.5$ has the same origin as the
''0.25 structure'' in conductance, a direct consequence of a sharp
singlet resonance. In Fig.~1(c) we show a schematic representation
of the non-linear regime with finite source-drain voltage. In this
case the double peak potential barrier is asymmetric and shallower,
giving rise to broader singlet and triplet resonances. The triplet
resonance can even become over-damped while the singlet becomes more
robust to temperature as it broadens. Hence a pronounced ''0.25 structure''
in the conductance is expected, surviving to higher source-drain voltage
than the triplet (0.7 structure). This is indeed seen in experiments.\cite{patel91,philmag98,picciotto04}
If Eq.~(\ref{eq:f}) at least qualitatively holds also in this non-linear
regime, a distinctive feature should appear in the Fano factor, as
presented in Fig.~2(${\textrm{a}}_{3}$,${\mathrm{b}}_{3}$,${\textrm{c}}_{3}$).

Thus far we have calculated conductance and Fano factor from singlet
and triplet resonances and found good semi-quantitative agreement
with experiment. In the regime of linear conductance and low temperature
Eq.~(\ref{eq:lbg}) and Eq.~(\ref{eq:f}) simplify, with $G$ and
$F$ determined by $T_{s}(\mu)$ taken at the Fermi energy. We can
then invert the procedure and use the experimentally determined conductance
and Fano factor to determine the transmission probabilities $T_{0}$
and $T_{1}$ using Eq.~(\ref{eq:lbg}) and Eq.~(\ref{eq:f}) in
this regime. Unitarity for $F$ requires that $F\leq1-G/G_{0}$ and
that $F$ is above some lower limit, $F_{min}(G)$, dotted line in
inset of Fig.~3(a). Unfortunately experimental values of $F$ rise
above this limit, possibly due to uncertain temperature corrections,
and therefore cannot be used to determine the probabilities unambiguously.
However, some estimates can be done. First we approximate $F(G)=\min(F_{I},1-G/G_{0}$),
where $F_{I}$ corresponds to the line connecting the experimental
points for $B=0$. Such a ''fit I'' is presented in the inset of
Fig.~3(a) (full line). Another choice is $F(G)=F_{II}$, where $F_{II}$
is some ''minimal assumption'' linear approximation for experimental
data at low $G$ and presented in Fig.~3(a) inset with a dashed line,
''fit II''. Line-shaded areas between the two choices correspond to
the experimentally undetermined regime. 

In Fig.~3(a) are presented singlet and triplet transmission probabilities
extracted from such $F(G)$ and the corresponding experimental values
for $G(\mu)$ for $T=$515~mK. In spite of the uncertainty in $F$,
the structure of both, $T_{0}$ and $T_{1}$ are relatively well determined
and remarkably similar to the theoretically predicted cases from Fig.~2,
with a much larger triplet transmission probability at lower energies,
where the singlet is just above resonance and only slowly approaches
unity. A small resonance in the singlet transmission probability corresponds
to the kink structure in $F$ at $G\sim0.5G_{0}$. However, more accurate
measurements of $F$ are necessary in order to reduce the error bars
in the estimates of singlet and triplet transmission probabilities.
In Fig.~3(b) these probabilities are extracted for the case of the
lower temperature $T=$273~mK. The most striking observation is the
more rapid increase of the singlet transmission probability, a possible
signature of Kondo behavior.\cite{cronenprl,meir02}

The experimentally measured Fano factor in a strong magnetic field
clearly suggests spin-up and spin-down structure of spin sub-channels.
Here the singlet--triplet concept is not relevant, $T_{\uparrow}$
and $T_{\downarrow}$ being the appropriate sub-channel division.
Eq.~(\ref{eq:f}) is therefore not valid in this limit. However,
the theoretical results for the conductance of near perfect quantum
wires in a magnetic field\cite{jphys00,rrj03} predict that, due to
the Zeeman sub-band splitting at finite magnetic field, only one spin
channel is open at lower energies and the conductance reaches $G\sim G_{0}$
only at higher energies, consistent with experiment. The corresponding
Fano factor then follows the unitarity limit for this case, $F\sim1-2G/G_{0}$
for $G<0.5G_{0}$ forming a bow with a maximum at $G\sim0.75G_{0}$
shown in Ref.~\onlinecite{roche}. This is qualitatively consistent
with the experimentally determined high field results for $F$, though
a more quantitative description of the transition between the two
regimes is still lacking.

To conclude, we have shown that anomalous structures in shot noise
Fano factor measurements can be understood within the framework of
the theory of 0.7 conductance anomalies in near perfect quantum wires.\cite{rrj00}
The analysis of temperature dependence indicates stronger \emph{singlet}-channel
temperature dependence, which could be related to the Kondo-like behavior
at lower temperature.\cite{cronenprl,meir02} High magnetic field
measurements are in qualitative agreement with the results of the
theory. Finally, our results for a weak asymmetric confining potential
in an otherwise perfect quantum wire predict that in finite source-drain
voltage measurements a strong structure in Fano factor should appear
for $G\lesssim0.5G_{0}$. This structure corresponds to the recently
measured 0.25 conductance anomalies.\cite{picciotto04} Additional
refined measurements of conductance and the Fano factor could more
precisely resolve the singlet and triplet transmission probabilities
and test the predictions of the theory based on the singlet--triplet
resonant scattering.

We thank M. B{\" u}ttiker for drawing our attention to the shot-noise
problem and for helpful suggestions. We thank also T. Rejec for comments
and B. Bu{\l}ka for stimulating discussions. Support is gratefully
acknowledged from the Ministry of Education and Science of Slovenia
under grant Pl-0044, the EU and the UK MoD.

\widetext

\begin{figure}[h]
\begin{center}\includegraphics[%
  clip,
  width=9cm,
  keepaspectratio]{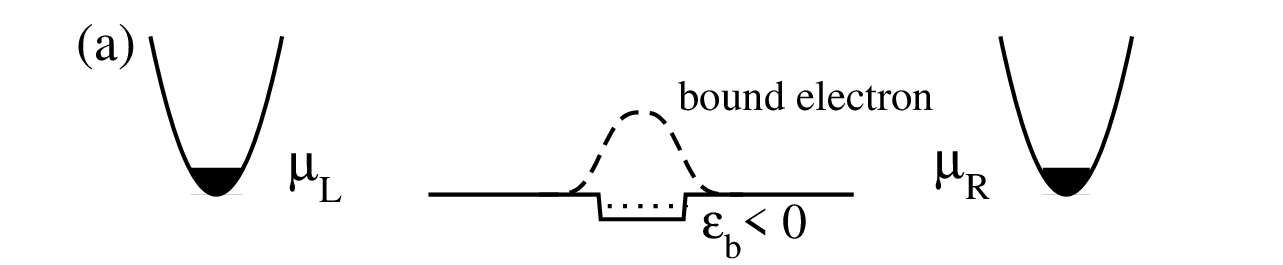}\end{center}

\begin{center}\includegraphics[%
  clip,
  width=9cm,
  keepaspectratio]{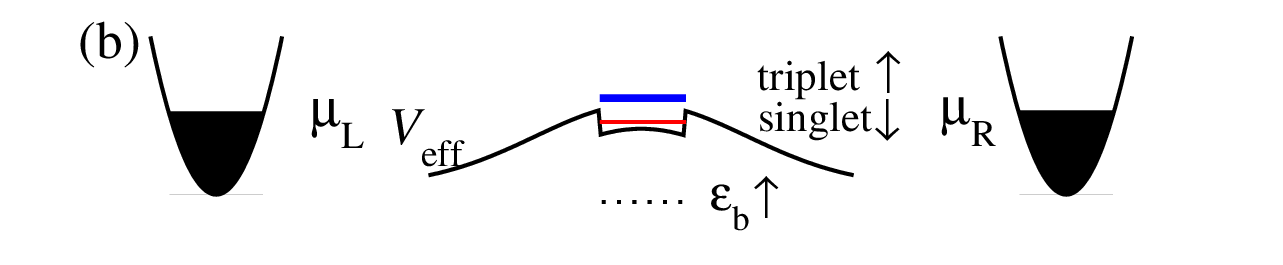}\end{center}

\begin{center}\vskip -.5cm\includegraphics[%
  clip,
  width=9cm,
  keepaspectratio]{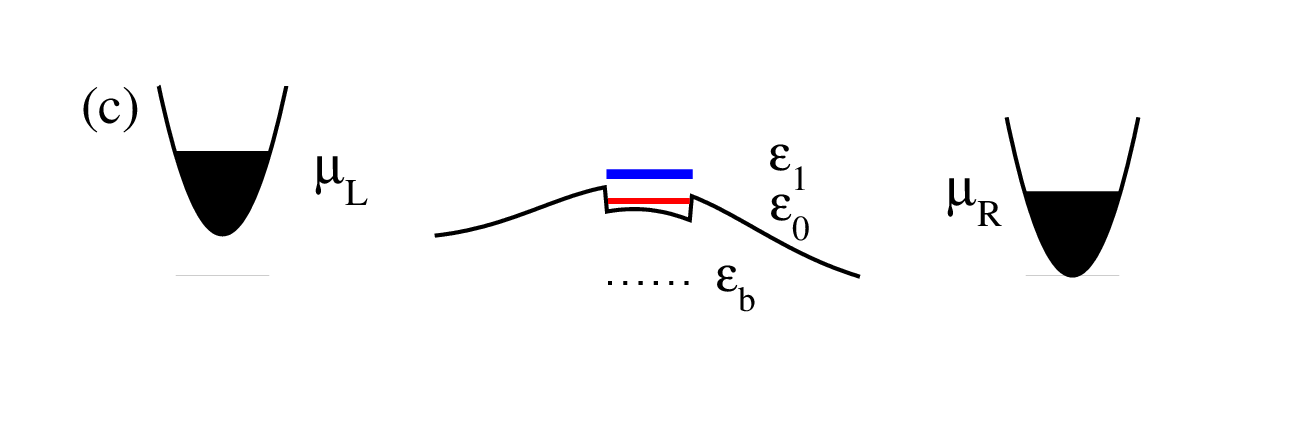}\end{center}

\caption{\label{cap:Fig1}(color online) (a) A weak negative potential fluctuation
binds one electron (electron density shown with dashed line) with
the energy below first channel minimum, $\epsilon_{b}<0$ (dotted
line) and chemical potential $\mu_{L}\sim\mu_{R}$. (b) Finite conduction
with electron energy, $\epsilon>0$, and linear regime, $\mu_{L}\gtrsim\mu_{R}$.
Two-electron scattering (quasi-bound) state is singlet or triplet.
$V_{eff}$ is an effective double barrier tunneling potential for
the scattered electron. (c) Larger source-drain voltage where the
triplet ($\epsilon_{1}$) resonance becomes over-damped with a broader,
more robust, singlet ($\epsilon_{0}$) becoming visible in transport.}
\end{figure}

\begin{figure}
\begin{center}\includegraphics[%
  clip,
  width=15cm,
  keepaspectratio]{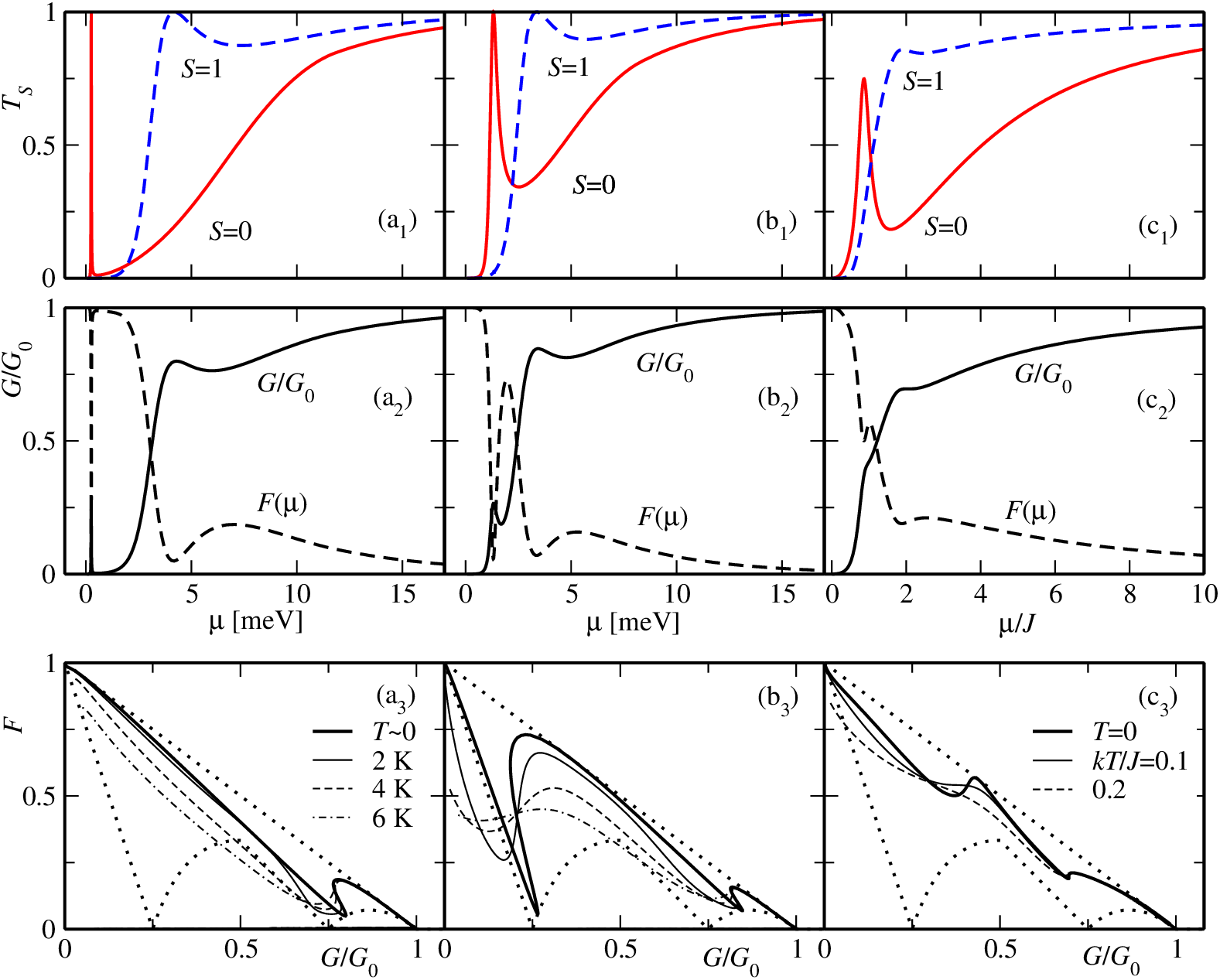}\end{center}

\caption{\label{cap:Fig2}(color online) Panels (${\textrm{a}}_{1}$) and
(${\textrm{b}}_{1}$): singlet and triplet transmission probabilities
as a function of Fermi energy $\mu$ measured relative to the bottom
of the first channel. Results for cylindrical quantum wires with a
symmetric bulge for all parameters as in Fig.~3(b) and Fig.~3(c)
of Ref.~\onlinecite{rrj00}. In panels (${\textrm{a}}_{2}$) and
(${\textrm{b}}_{2}$) the conductance (full lines) and the Fano factor
are presented. Panels (${\textrm{a}}_{3}$) and (${\textrm{b}}_{3}$):
Fano factor as a function of $G$ (full line) and unitarity limits
(dotted line). (${\textrm{c}}_{1}$) transmission probabilities as
would arise, e.g., for a left-right asymmetric confining potential
(in resonance $T_{S}<1$). The corresponding energy dependence of
the Fano factor and conductance, (${\textrm{c}}_{2}$), and Fano factor
as a function of conductance, (${\textrm{c}}_{3}$). Energy and temperature
scale is here in the units of single-triplet energy difference $J$.}
\end{figure}

\begin{figure}
\begin{center}\includegraphics[%
  clip,
  width=12cm,
  keepaspectratio]{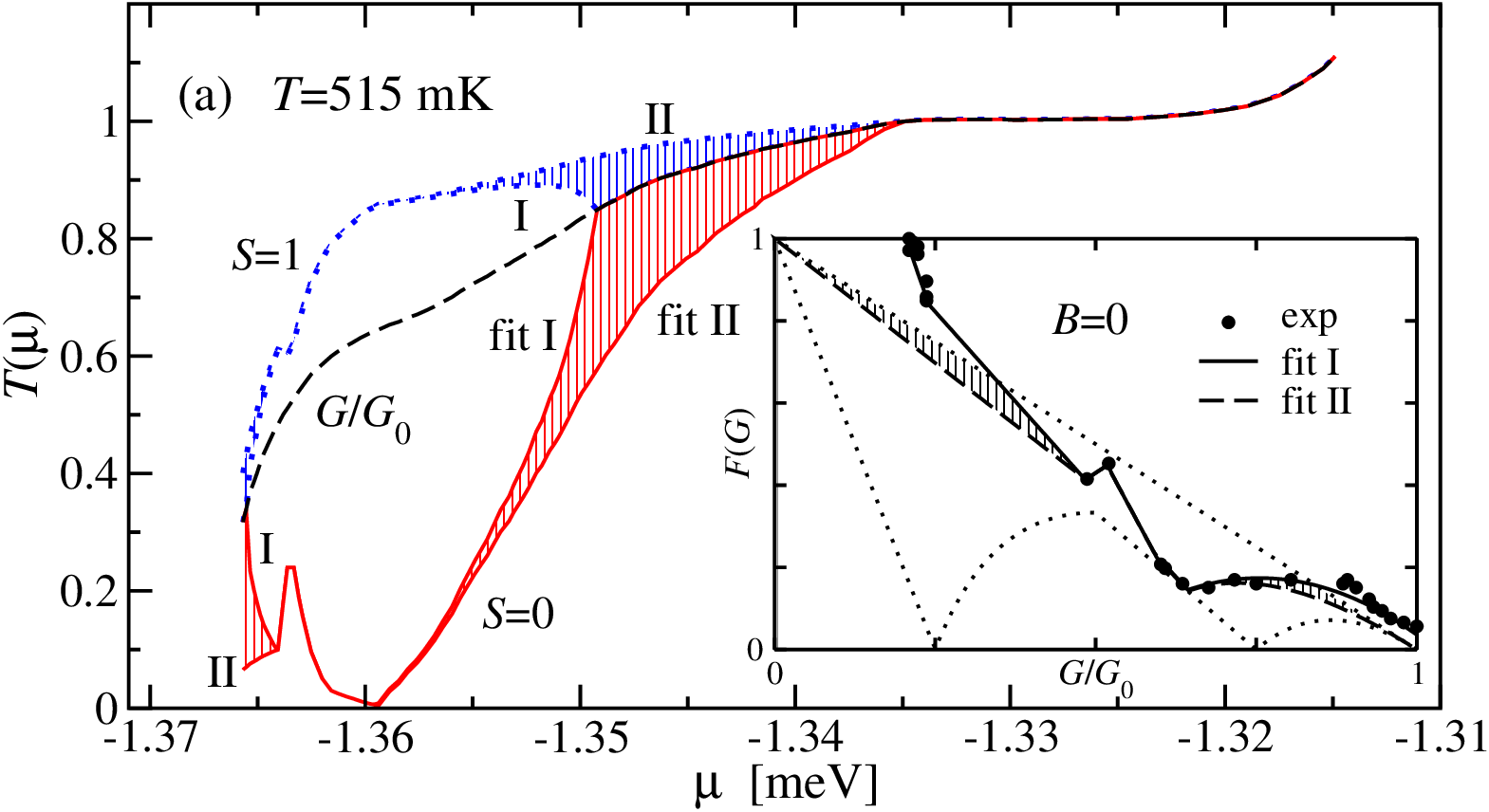}\end{center}

\begin{center}\includegraphics[%
  clip,
  width=12cm,
  keepaspectratio]{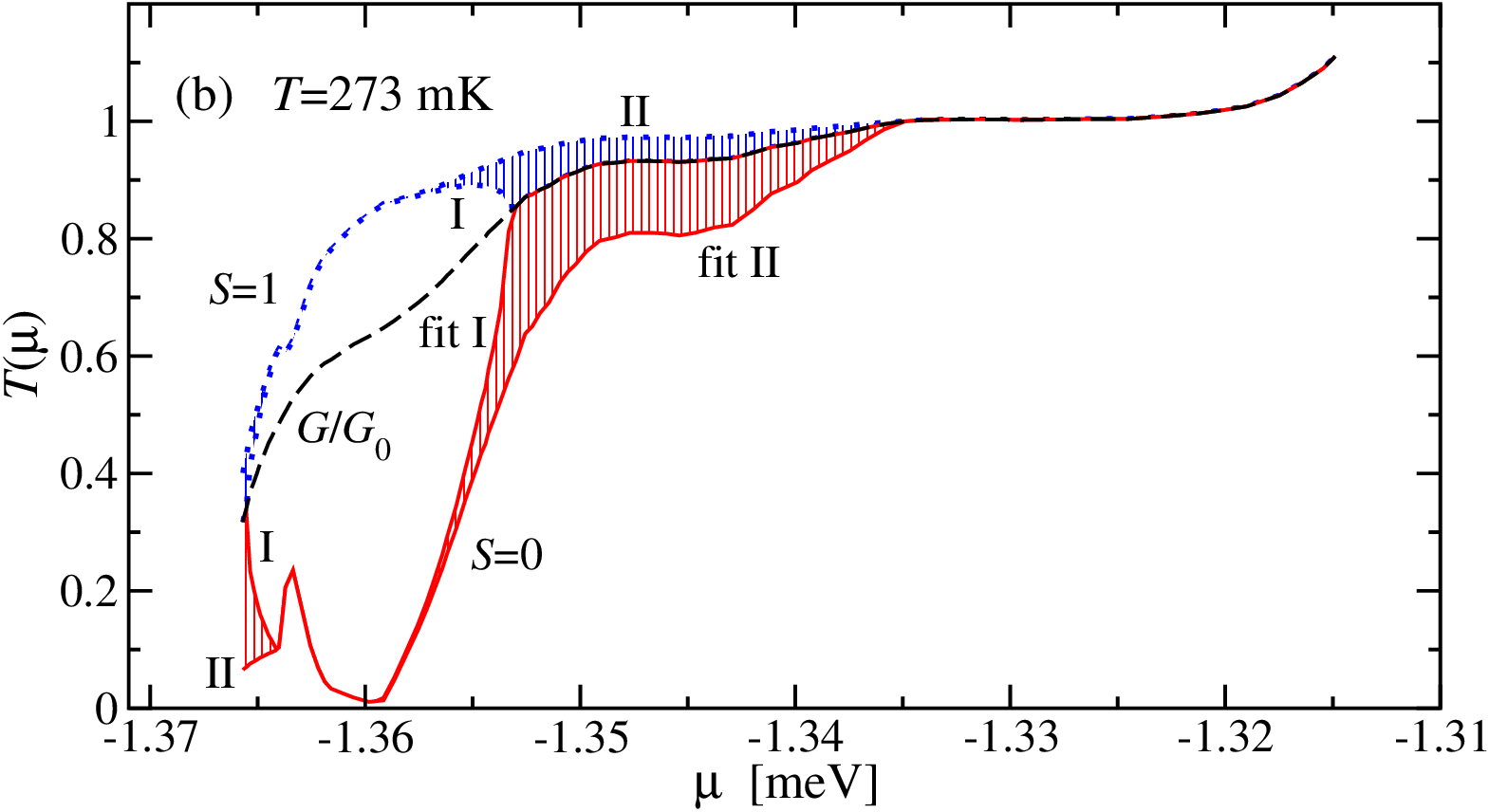}\end{center}

\caption{\label{cap:Fig3}(color online) (a) Inset: experimental values from
Ref.~\onlinecite{roche} for $F(G)$ for $B=0$ (bullets) and two
different interpolating (fitting) forms of $F$. Main figure: conductance
for $T=515$ mK and $B=0$ (dashed line), singlet (full line) and
triplet (dotted) transmission probabilities extracted from experimental
values of $G(\mu)$ vs $\mu$ and using two interpolating forms for
$F$. Shaded area corresponds to the shaded area in the inset. (b)
As in (a), for $T=273$ mK.}
\end{figure}

\end{document}